\documentclass[manuscript]{aastex}
\usepackage{graphicx}
\usepackage{natbib}
\usepackage{upgreek}
\usepackage{color}

\newcommand{\kep}{$K_{ep}$}
\newcommand{\nh}{\bar{n}_{\rm H}}

\newcommand{\hi}{H$\,\scriptstyle{\rm I}$~}

\slugcomment{To be resubmitted to ApJ: v23,  \today}

\shorttitle{\emph{Fermi} LAT Observations of the Monoceros Loop SNR}
\shortauthors{\emph{Fermi} LAT collaboration}

\begin{document}

%\title{\emph{Fermi} Large Area Telescope Observations of the Monoceros Loop Supernova Remnant}
\title{\emph{Fermi} LAT Study of Gamma-Ray Emission
in the Direction of the Monoceros Loop Supernova Remnant}

\author{
H.~Katagiri\altaffilmark{1,2},
S.~Sugiyama\altaffilmark{1},
M.~Ackermann\altaffilmark{3}, 
J.~Ballet\altaffilmark{4},
J.~M.~Casandjian\altaffilmark{4}, 
Y.~Hanabata\altaffilmark{5}
J.W.~Hewitt\altaffilmark{6}, 
M.~Kerr\altaffilmark{7}, 
H.~Kubo\altaffilmark{8},
M.~Lemoine-Goumard\altaffilmark{9},
P.~S.~Ray\altaffilmark{10} 
%L.~Tibaldo\altaffilmark{11}
}
\altaffiltext{1}{College of Science, Ibaraki University, 2-1-1 Bunkyo, Mito, Ibaraki 310-8512, Japan}
\altaffiltext{2}{Corresponding author: H.~Katagiri and S.~Sugiyama, hideaki.katagiri.sci@vc.ibaraki.ac.jp}
\altaffiltext{3}{Deutsches Elektronen Synchrotron DESY, D-15738 Zeuthen, Germany}
\altaffiltext{4}{Laboratoire AIM, CEA-IRFU/CNRS/Universit\'e Paris Diderot, Service d'Astrophysique, CEA Saclay, F-91191 Gif sur Yvette, France}
\altaffiltext{5}{Institute for Cosmic-Ray Research, University of Tokyo, 5-1-5 Kashiwanoha, Kashiwa, Chiba 277-8582, Japan}
\altaffiltext{6}{University of North Florida, Department of Physics, 1 UNF Drive, Jacksonville, FL 32224 , USA}
\altaffiltext{7}{CSIRO Astronomy and Space Science, Australia Telescope National Facility, Epping NSW 1710, Australia}
\altaffiltext{8}{Department of Physics, Graduate School of Science, Kyoto University, Kyoto, Japan}
\altaffiltext{9}{Centre d'\'Etudes Nucl\'eaires de Bordeaux Gradignan, IN2P3/CNRS, Universit\'e Bordeaux 1, BP120, F-33175 Gradignan Cedex, France}
\altaffiltext{10}{Space Science Division, Naval Research Laboratory, Washington, DC 20375-5352, USA}
%\altaffiltext{11}{Max-Planck-Institut f\"ur Kernphysik, D-69029 Heidelberg, Germany}

%\author{The Fermi LAT Collaboration}
%\affil{Contact Authors: Hideaki Katagiri, Luigi Tibaldo \\
%Participants: Marshall Roth,
%Omar Tibolla, Y. Uchiyama, Jean Ballet,  R. Yamazaki \\
%Internal reviewers: Francesco Giordano, Troy Porter
%} 
%\email{katagiri@hep01.hepl.hiroshima-u.ac.jp, luigi.tibaldo@pd.infn.it} 

\begin{abstract}
We present an analysis of the gamma-ray measurements by the Large Area Telescope onboard
the \textit{Fermi Gamma-ray Space Telescope} in the region of
the supernova remnant~(SNR) Monoceros Loop~(G205.5$+$0.5). 
%
% david changed this sentence:
%We find a significant detection at $\sim 21 \sigma$ of gamma-ray emission 
%associated with the SNR in the energy band 0.2--100~GeV.
%We detect significant gamma-ray emission associated with the SNR in the energy band 0.2--300~GeV. 
The brightest gamma-ray peak is spatially correlated with the Rosette Nebula, which is a molecular cloud complex adjacent to the southeast edge of the SNR. After subtraction of this emission by spatial modeling, the gamma-ray emission from the SNR emerges, which is extended and fit by a Gaussian spatial template.
The gamma-ray spectra are significantly better reproduced by a curved shape than a simple power law. The luminosities between 0.2--300~GeV are 
$\sim$~$4 \times 10^{34}$~erg~s$^{-1}$ for the SNR and $\sim$~$3 \times 10^{34}$~erg~s$^{-1}$ for the Rosette Nebula, respectively.
%$\sim$~$4.1 \times 10^{34}$~erg~s$^{-1}$ for the SNR and $\sim$~$2.7 \times 10^{34}$~erg~s$^{-1}$ for the Rosette Nebula, respectively.
%Given the association between radio continuum emission and gamma-ray emission around the SNR,
%We argue that gamma rays originate in interactions between particles accelerated in the SNR and interstellar gas or radiation fields.
We argue that the gamma rays likely originate from the interactions of particles accelerated in the SNR.
The decay of neutral pions produced in nucleon-nucleon interactions of accelerated hadrons with interstellar gas provides a reasonable explanation 
for the gamma-ray emission of both the Rosette Nebula and the Monoceros SNR.

\end{abstract}


\section{Introduction}
The shock waves of supernovae accelerate particles to very high energies through the mechanism of diffusive shock acceleration~\citep[e.g.,][]{blandford87}.
However, the processes of acceleration, release from the shock region, and diffusion in the interstellar medium of such particles are not well understood.
Gamma-ray observations in the GeV domain are a powerful probe of these mechanisms.
The Large Area Telescope~(LAT) on board the \textit{Fermi Gamma-ray Space Telescope} has detected GeV gamma rays from several SNRs~\citep[e.g., ][and references therein]{SNRref, SNRCat}. 

The Monoceros Loop~(G205.5$+$0.5) is a well-studied middle-aged SNR. It has a large diameter~($\sim$~$3\fdg8$) which allows detailed morphological studies in high-energy gamma rays since the LAT has a comparable point-spread function~(PSF) above a few hundred MeV~(the 68\% containment angle above 1~GeV is smaller than 1$^\circ$). The radio emission of the SNR has a non-thermal spectrum~\citep[e.g.][]{RadioSpectrum}, indicating the existence of high-energy electrons. A young stellar cluster and molecular cloud complex, the Rosette Nebula, is located at the edge of the southern shell of the SNR. 
% rosette = (NGC~2244)
%The nebula contains a cluster of several ionizing early O-type stars~(NGC~2264) known to be losing appreciable amounts of mass to the ambient interstellar medium~\citep{rosette}. 
%The Monoceros Loop is suggested to be interacting with the Rosette Nebula. 
The H$\alpha$ line widths in the SNR ridge overlapping with the Rosette region
are larger than near the center of the Rosette Nebula~\citep{halpha}, suggesting that the SNR is interacting with the Rosette Nebula.          
%In the SNR ridge overlapping with the Rosette region, \cite{halpha} found broadened line widths of H109$\alpha$ and much enhanced H$\alpha$ emission compared to the average in the Rosette Nebula, suggesting that the SNR is interacting with the Rosette Nebula.
\cite{distance1} obtained distances
of $1.6~\mathrm{kpc}$ for stars associated with the Rosette Nebula
from main sequence fitting. 
\cite{distance2} argued that the distance of the SNR is $1.6~\mathrm{kpc}$ 
%by summarizing all the available evidences together with 
based on his decameter wavelength observations of the absorption of nonthermal emission from the SNR by the Rosette Nebula.
% This was further supported by the decameter wavelength observations of the absorption of nonthermal emission from the SNR by the Rosette Nebula~\citep{distance2}.
In this paper, we adopt this distance of 1.6~kpc for both objects.
The age was estimated to be $\sim$~$3 \times 10^4$~yr based on the X-ray data and an SNR model~\citep{age}.

In Monoceros, a very-high-energy~(VHE) gamma-ray source, HESS~J0632$+$057, was first discovered at TeV energies by the High Energy Stereoscopic System~\citep[H.E.S.S.;][]{HESS}, located close to the rim of the Monoceros SNR.
It appears to be point-like within experimental resolution; the limit on the size of the emission region was given as $2\arcmin$~(95\% confidence level). 
Detection of variability in the VHE gamma-ray and X-ray fluxes supports interpretation of the object as a gamma-ray emitting binary~\citep{VERITAS}, indicating that the bulk of VHE gamma rays do not come from high-energy particles accelerated by the SNR.
No significant emission from the location of HESS~J0632$+$057 was detected in the 0.1-100 GeV energy range integrating over 3.5 yr of \emph{Fermi} LAT data~\citep{Fermi-HESSJ0632}.
Also, an unidentified high-energy gamma-ray source, 3EG~J0634$+$0521 has been detected using the EGRET data~\citep{3eg}.
% No AGILE source
%However, morphological study such as association with molecular clouds requires higher photon statistics with better angular resolution.
However, morphological studies which could associate the gamma-ray emission with molecular clouds require higher photon statistics with better angular resolution.

Seven LAT point-like sources positionally associated with the Monoceros Loop SNR~(2FGL~J0631.6$+$0640, 2FGL~J0633.7$+$0633, 2FGL~J0636.0$+$0554, and 2FGL~J0637.8$+$0737) and the Rosette Nebula~(2FGL~J0631.7$+$0428, 2FGL~J0634.3$+$0356c, and 2FGL~J0637.0$+$0416c) are listed in the 2FGL catalog~\citep{2yrCatalog}.
However, 
the two sources~(2FGL~J0634.3$+$0356c and 2FGL~J0637.0$+$0416c) associated with the Rosette Nebula were classified as `c' sources that require caution in interpreting or analyzing. 
In addition, the extended emission around the SNR was reported in the first \emph{Fermi} LAT SNR catalog~\citep{SNRCat}, which was modeled as a uniform disk with the radius of 2\fdg3. 
In order to understand the emission and its mechanism more deeply,
a detailed analysis and further discussion are required.

In this paper, we report a detailed study of the emission in the direction of 
 the Monoceros Loop by using the \emph{Fermi} LAT data.
We have analyzed the 67-month LAT data by using the 2FGL catalog. {\bf Observations and data selection are briefly described in Section~\ref{sec:obs}. The analysis procedure and results described in Section~\ref{sec:ana} include a study of the morphology and spectrum of the emission associated with the Monoceros Loop and the Rosette Nebula. Finally, we present our results in Section~\ref{sec:discuss} and conclusions in Section~\ref{sec:conclusion}.
The data selection, analysis procedure, and the modeling of gamma-ray emission are based on the previous studies of the Cygnus Loop by \cite{CygnusLoop} and SNR HB~3 by \cite{HB3}}.

\section{OBSERVATIONS AND DATA SELECTION}
\label{sec:obs}
{\bf The main instrument on \emph{Fermi} is the LAT which detects gamma rays from $\sim$~20~MeV to $>$~300~GeV\footnote{\bf Only events with energies $>$~0.2~GeV are used in this analysis.}. 
The LAT is an electron-positron pair production telescope, using tungsten foil converters and silicon microstrip detectors and a hodoscopic cesium iodide calorimeter to measure the arrival directions and energies of incoming gamma rays.}
They are surrounded by 89 segmented plastic scintillators that serve as an anticoincidence detector to reject events originating from charged particles.
Detailed information about the instrument can be found in \cite{Atwood09}, the on-orbit calibration is described in \cite{onorbit_cal}, and a summary of event classification strategies and instrument performance is given in \cite{IRFs}.
The LAT has a larger field of view~($\sim$~2.4~sr), a larger effective area~($\sim$~8000~cm$^2$ for $>$1~GeV on-axis peak effective area) and improved PSF in comparison to previous high-energy gamma-ray telescopes.
%point-spread function~(PSF; the 68\% containment angle $>1$~GeV is smaller than 1$^\circ$). 
% angular resolution
% {\bf The tracker of the LAT is divided into front and back sections. The front section (first 12 planes) has thin converters to improve the point-spread function~(PSF), while the back section (four planes after the front section) has thicker converters to enlarge the effective area. The angular resolution of the front events is a factor of 2 better than that of the back events at 1~GeV.}

{\bf We analyzed events toward the Monoceros Loop recorded from the start of science operations on 2008 August 4 until 2014 January 29. The LAT operated in a nearly continuous sky survey mode, to obtain a total exposure of $\sim$~1.5~$\times$~10$^{11}$~cm$^2$~s~(at 1~GeV). In this observing mode approximately uniform coverage of the entire sky is obtained every 2 orbits~($\sim$~3~hr).}

We used the standard LAT analysis software, the \emph{ScienceTools} version v9r32,
publicly available from the \emph{Fermi} Science Support Center (FSSC)\footnote{Software and
documentation of the \emph{Fermi} \emph{ScienceTools} are distributed by the \emph{Fermi} Science
Support Center at \url{http://fermi.gsfc.nasa.gov/ssc}.}.
We use events classified as \emph{P7SOURCE} that have been reprocessed with an updated instrument calibration~\citep{P7REP}. Only events that have a reconstructed zenith angle less than 100$^\circ$ were used in order to minimize the contamination from Earth-limb gamma-ray emission.
Furthermore, only time intervals when the center of the LAT field of view is within 52$^\circ$ of the local zenith are accepted to further reduce the contamination by Earth's atmospheric emission. The Instrument response functions~(IRFs) that correspond to this dataset are \texttt{P7REP\_SOURCE\_V15}~(publicly available via the FSSC) {\bf throughout this work.}
%We used the post-launch instrument response functions~(IRFs) \texttt{P7REP\_SOURCE\_V15}~\citep{IRFs} and applied the following event selection criteria: 
%1) events should be classified as so-called reprocessed Pass 7 \emph{Source} class \citep{Atwood09}, 2) events have a reconstructed zenith angle less than 100$^\circ$, to minimize the contamination from Earth-limb gamma-ray emission,
% and 3) only time intervals when the center of the LAT field of view is within 52$^\circ$ of the local zenith are accepted to further reduce the contamination by Earth's atmospheric emission.

{\bf Times when the LAT detected a gamma-ray burst (GRB) or nova were eliminated from the dataset. 
The transients located within 15$^\circ$ of the Monoceros Loop were GRB~130504C~\citep{GRB} and Nova Mon 2012~\citep{nova}, corresponding to 56416.97797--56417.00390 and  56099.00000--56109.00000 in Modified Julian Day, respectively.}

\begin{figure}
%\epsscale{.80}
%\plotone{f1.pdf}
%\plotone{figs/lc.eps}
%\plotone{figs/lc_v3.eps}
\plotone{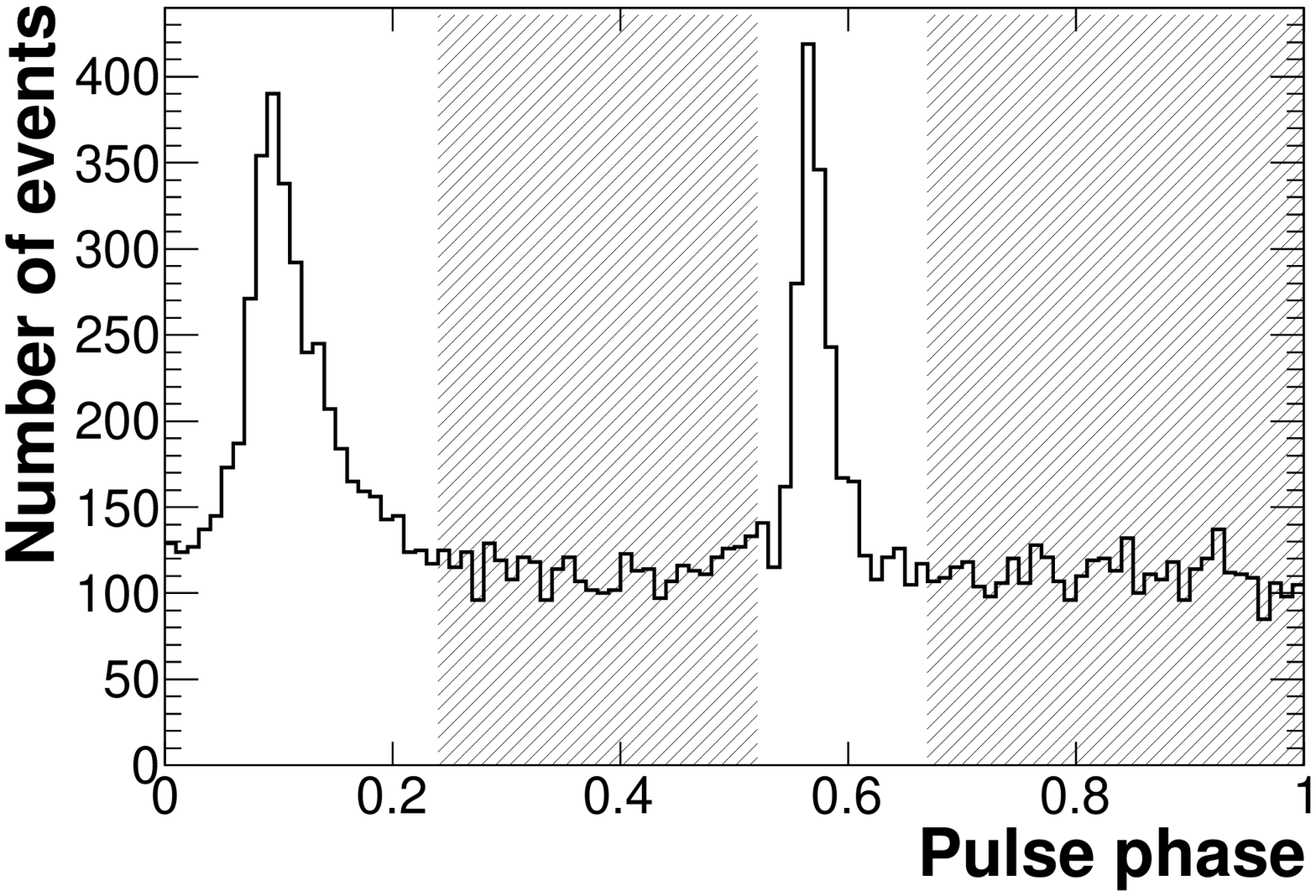}
\caption{
Pulse profile of PSR~J0633$+$0632 using \emph{Fermi} LAT data
for photon energies 0.2--300~GeV. The off-pulse phase range used in this analysis is shown by shaded regions.
\label{fig:pulse}
}
\end{figure}

The region around the SNR is dominated by the gamma-ray emission of PSR~J0633$+$0632 (2FGL~J0633.7$+$0633). The pulse profile in the 0.2--300~GeV energy range analyzed in this paper is shown in Figure~\ref{fig:pulse}, where the events are within 1$^\circ$ of the pulsar position.
Using a timing solution modeling the effects of spin-down and timing noise~\citep{kerr15}, we assigned rotational phase to each photon 
using the \emph{Fermi} plug-in of the \texttt{TEMPO2} software package~\citep{TEMPO2}.
%A phase was associated with each event using the Fermi plug-in of the \texttt{TEMPO2} software package~\citep{TEMPO2}.
We only used the events during the off-pulse phases of PSR~J0633$+$0632, corresponding to phases of 0.24--0.52 and 0.67--1.00
as adopted in \cite{2PC}.
%{\bf as adopted in \cite{Fermi-HESSJ0632}}. 
%In addition to these event selections,
%On the following analysis in this paper, 
We restricted the energy range to $>~0.2$~GeV to avoid possible large systematics due to the rapidly varying effective area and much broader PSF at lower energies.

\section{ANALYSIS AND RESULTS}
\label{sec:ana}
\subsection{General settings}
\label{subsec:settings}

{\bf The morphology and spectrum of gamma-ray emission from the Monoceros Loop and Rosette Nebula were determined using a} binned likelihood analysis based on Poisson statistics\footnote{As implemented in the publicly available \emph{Fermi} \emph{Science Tools}.
The documentation concerning the analysis tools and the likelihood fitting procedure is available from
\url{http://fermi.gsfc.nasa.gov/ssc/data/analysis/documentation/Cicerone/}.}~\citep[see, e.g.,][]{Mattox96}.
The likelihood is the product of the probabilities of the observed gamma-ray counts within each spatial and spectral bin for a specified model.
% The probability density function for the likelihood analysis
{\bf The gamma-ray emission model used here included all sources detected in the 2FGL catalog within 20$^\circ$ of the SNR. We also included the standard LAT diffuse background model~\citep{diffusemodel}, \texttt{gll\_iem\_v05\_rev1.fit} that results from cosmic-ray~(CR) interactions with the interstellar medium and radiation fields and an isotropic component to represent extragalactic gamma rays and charged particle background using a tabulated spectrum~(\texttt{iso\_source\_v05.txt}). Both diffuse models are available from the FSSC.} We fit all spectral parameters of the 2FGL sources spatially associated with the SNR~(3 sources within the SNR~(Group S), 3 sources within the Rosette Nebula~(Group R) and PSR~J0633$+$0632), the Galactic diffuse emission and the isotropic component, while the integral fluxes of the other point sources {\bf are left as free parameters and the spectral indices are fixed to the values reported in 2FGL.}

\begin{figure}
%\plotone{figs/map0/cmap_paper.eps}
\plotone{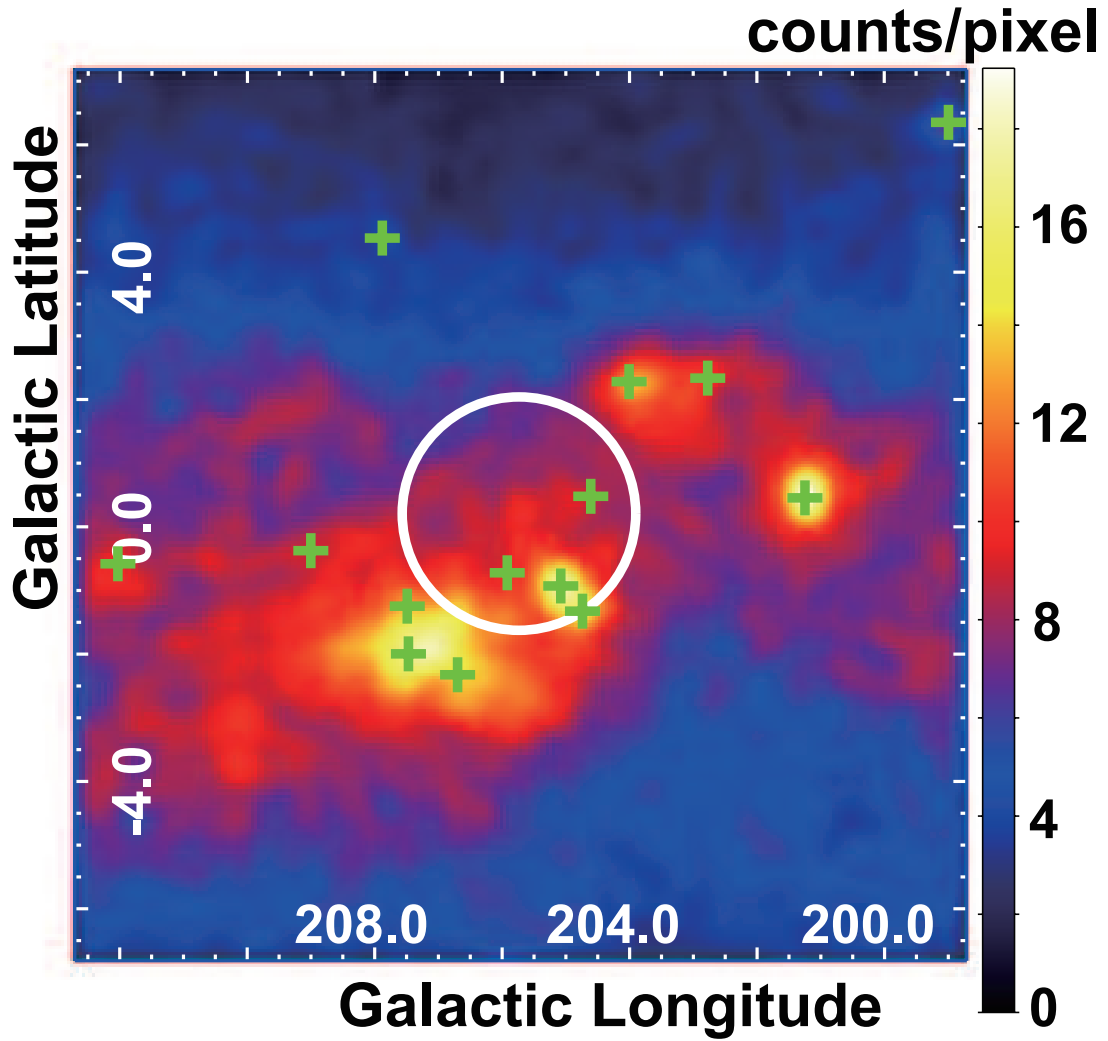}
\caption{
\small
LAT counts map of the region of interest in the 0.5--10~GeV energy range
produced using the events during the off-pulse phases of PSR~J0633$+$0632. The details of the off-pulse analysis are shown in the text.
The counts map is binned using a grid of $0\fdg1$ and smoothed with a Gaussian kernel of $\sigma$ $=$ $0\fdg2$.
The white circle shows the position of the Monoceros Loop. 
Crosses indicate the positions of gamma-ray sources listed in the 2FGL
catalog~\citep{2yrCatalog}.
\label{fig:cmap}
}
\end{figure}

The analyses were performed {\bf within a 14$^\circ \times$14$^\circ$ square region using $0\fdg1$ pixels.}
The energy range for likelihood analysis is divided into 40 logarithmically-spaced energy bins from 0.2~GeV to 300~GeV.
Figure~\ref{fig:cmap} shows the counts map in the region of interest.
We centered the region on the center of the SNR: 
%(R.A., Dec.)~$=$~(6${}^h$39${}^m$00${}^s$, 6${}^\circ$30$'$00$''$)~(J2000).
(R.A., Dec.)~$=$~(99\fdg75,6\fdg50)~(J2000).
% coordinate converter
% J2000 B1950 Galactic Ecliptic
%RA Dec RA Dec L B Lon Lat
%06 39 00.00 06 30 00.0 06 36 18.73 06 32 43.9 205 43 56.5 00 12 33.4 100 06 43.9 -16 35 51.7
%99.750000 6.500000 99.078023 6.545531 205.732350 0.209279 100.112206 -16.597690

\subsection{Morphological analysis}
\label{subsec:spatial}

\begin{figure}
%\epsscale{0.9}
%\plottwo{f2b.pdf}{f2c.pdf}
% \leavevmode
% \includegraphics[scale=.26]{f2a.eps}
%\epsscale{0.4}
%\plotone{figs/map1/residual_Monoceros_v2.eps}
%\plotone{figs/map_all/map_co_snr.eps}
%\plotone{figs/map2_v3/map2.eps}
%\plotone{figs/map2_v3/map2_150824.eps}
%\epsscale{1.5}
\plotone{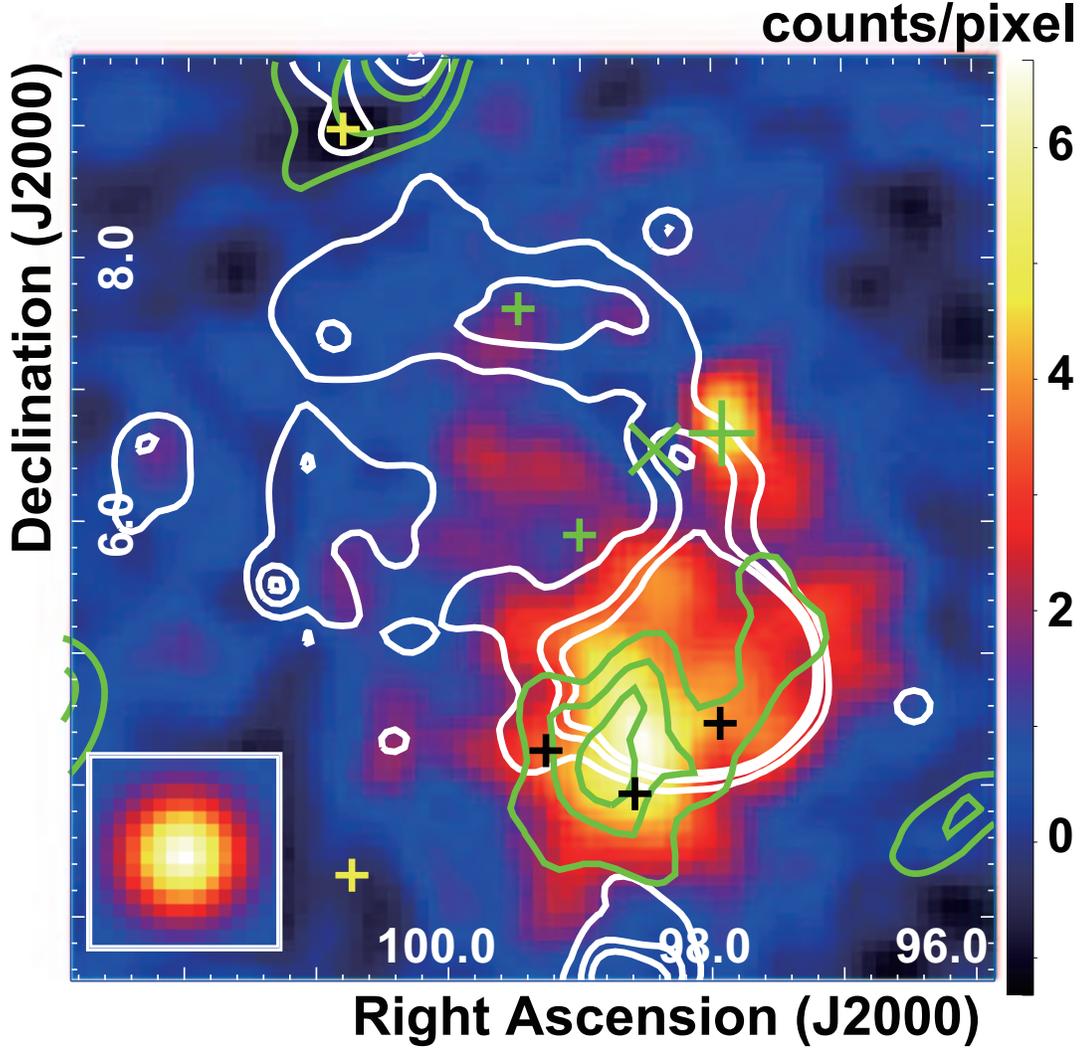}
\caption{
\small
Background-subtracted LAT counts map in the 0.5--10~GeV energy range.
The six LAT point sources associated with the Monoceros Loop and the Rosette Nebula except for PSR~J0633$+$0632 are not included in the background model. 
The counts map is binned using a grid of $0\fdg1$ and smoothed with a Gaussian kernel of $\sigma$ $=$ $0\fdg2$.
The inset of the figure shows the simulated LAT PSF with a photon index
of 2.5 in the same energy range, adopting the same smoothing.
%Note that all along the paper the analysis is conducted on unsmoothed data taking into account the instrument PSF in the likelihood analysis.
%Negative residuals are shown to gauge the quality of the subtraction of the background emission.
Crosses indicate the positions of gamma-ray sources listed in the 2FGL
catalog~\citep{2yrCatalog}.
The green, black, and yellow crosses are for the SNR, the Rosette Nebula, and the others, respectively.
The large green cross shows the position of 2FGL~J0631.6$+$0640.
The green x-mark indicates the position of PSR~J0633$+$0632 (2FGL~J0633.7$+$0633).
Green contours correspond to images at $^{12}$CO~($J=1\rightarrow0$) line intensities~\citep{CO}; 
 contours are at 5, 10, and 15~K~km~s$^{-1}$.
% The CO image was smoothed using a Gaussian kernel with $\sigma$ $=$~0.$^\circ$2. => not smoothed
White contours are Effelsberg 21~cm radio continuum~\citep{effelsberg}; contours are at 0.4, 0.8, and 1.2~K.
% The radio image was smoothed using a Gaussian kernel with $\sigma$ $=$~0.$^\circ$2; not smoothed
\label{fig:subtract_cmap}
}
\end{figure}

{\bf For our morphological study we only used events with energies greater than} 0.5~GeV~(compared to the 0.2~GeV used in our spectral analysis) to take advantage of the narrower PSF at higher energies.
Figure~\ref{fig:subtract_cmap} shows the counts map in a 7$^\circ~\times$~7$^\circ$ region centered on the Monoceros Loop, 
after subtracting the background: the Galactic emission, the isotropic component, and the 2FGL point sources except for the six 2FGL sources in Group S and R, the parameters of which were the best-fit ones obtained by the likelihood analysis where the emission associated with the SNR and the Rosette Nebula are modeled as the six sources~(Model 1 in Table~\ref{tab:likeratio1}).
The CO contours overlaid on the map correspond to line intensity integrated
over velocities of $0$~km~s$^{-1}$~$<V<20$~km~s$^{-1}$ with respect to the local standard of rest,
encompassing the velocity of 14~km~s$^{-1}$ corresponding to the distance from the Earth~(1.6~kpc assuming the IAU-recommended values $R_0 = 8.5$~kpc and $\Theta_0 = 220$~km~s$^{-1}$).
% Contours from radio observations are overlaid on the counts map. One of the contours corresponds to the Effelsberg 21~cm radio continuum~\citep{effelsberg}. The other contour corresponds to the CO 2.6~mm line~\citep{CO}, where the CO line intensities were integrated for velocities of $0$~km~s$^{-1}$~$<V<20$~km~s$^{-1}$ with respect to the local standard of rest (the velocity corresponding to the distance from the Earth~($1.6~\mathrm{kpc}$) is $\sim$~14~km~s$^{-1}$). 
The correlation between gamma rays and the CO line emission around the Rosette Nebula is evident.
We note that the LAT standard diffuse model includes this CO emission~\citep{diffusemodel}. Thus
the residual excess of the gamma-ray emission indicates that the CR density in this region is enhanced relative to the surrounding region.
The emission north of the CO region appears point-like and is consistent with the position of the source 2FGL~J0631.6$+$0640.
%Also, the point-like emission to the north of the CO emission is positionally correlated with 2FGL~J0631.6$+$0640.

%To quantitatively evaluate the correlation with emission at the CO line emission, 
To evaluate the correlation between the gamma-ray and CO line emission quantitatively, 
we fit LAT emission with a spatial template based on the CO line emission for the Rosette Nebula instead of the three 2FGL sources in Group R.
%The shape of CO emission was extracted only around the Rosette Nebula. 
We restricted the spatial template to a 2\fdg13 radius about the central cloud (R.A., Dec.)~$=$~(98\fdg41,4\fdg81)~(J2000).
Since the edge of the CO emission region is unclear 
due to statistical noise in the CO spectral measurements,
we introduced the CO intensity threshold used to create the spatial template as an additional free parameter in the fit. The spectral model was
 assumed to be a power-law function.
The resulting maximum likelihood values with respect to the maximum likelihood for the null hypothesis~(no source component associated with the Rosette Nebula and the SNR other than PSR~J0633$+$0632) are summarized in Table~\ref{tab:likeratio1}.
The test statistic~(TS) value~\citep[e.g.][]{Mattox96}
%, i.e. $-2\ln({\rm likelihood~ratio})$, 
for the CO image~(Model 2 in Table~\ref{tab:likeratio1}) is significantly larger than for {\bf all 3 individual point sources of Group R}~(Model 1). 
The threshold value of the CO intensity to maximize the likelihood value is 0~K$\cdot$km~s$^{-1}$ where the parameter varies from 0~K$\cdot$km~s$^{-1}$ to 4~K$\cdot$km~s$^{-1}$.

\begin{figure}
%\plottwo{f2b.pdf}{f2c.pdf}
% \leavevmode
% \includegraphics[scale=.26]{f2a.eps}
%\epsscale{0.4}
%\plotone{figs/map2/Monoceros_only3.eps}
%\plotone{figs/map2_v2/monoceros.eps}
%\plotone{figs/map_all/map_snr.eps}
%\plotone{figs/map3/map3.eps}
\epsscale{1}
%\plotone{figs/map3/map3_151228.eps}
\plotone{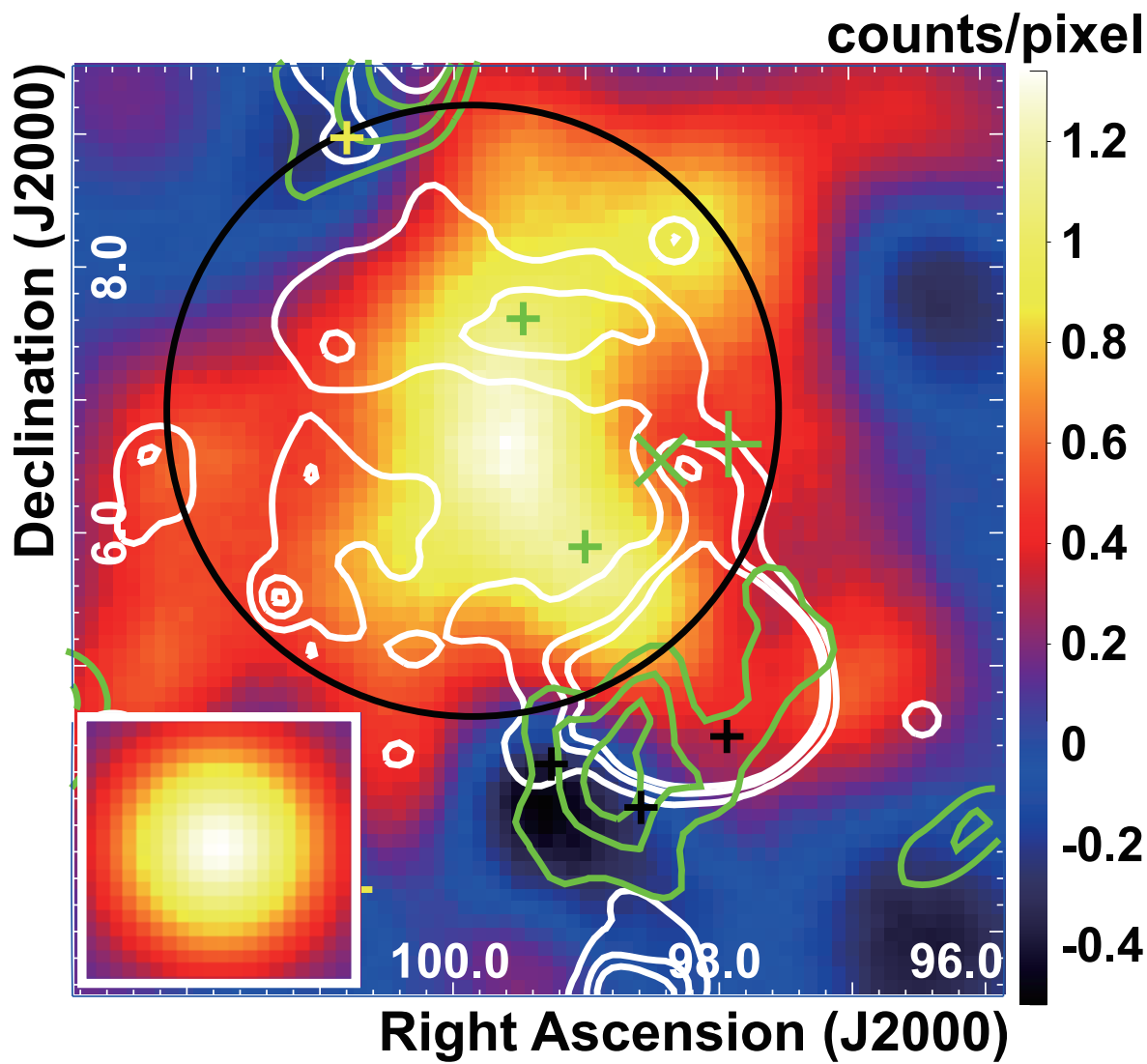}
\caption{
Background-subtracted LAT counts map in the 0.5--10~GeV energy range.
The emission predicted from the CO spatial template for the Rosette Nebula is subtracted, whereas the two 2FGL point sources associated with the SNR
(2FGL~J0636.0$+$0554, 2FGL~J0637.8$+$0737) are not included in the background model.
The counts map is binned using a grid of $0\fdg1$ and smoothed with a Gaussian kernel of $\sigma$ $=$ $0\fdg5$.
The black circle shows the best-fit Gaussian spatial model~(1-$\sigma$ radius).
The details of the overlays are described in the caption of Figure~\ref{fig:subtract_cmap}.
\label{fig:subtract_cmap2}
}
\end{figure}

We further characterized the morphology of gamma-ray emission associated with the Monoceros Loop. 
Figure~\ref{fig:subtract_cmap2} shows the counts map in a 7$^\circ~\times$~7$^\circ$ region centered on the SNR, after subtracting the background
and the emission from the Rosette Nebula. The background here consists of the Galactic emission, the isotropic component, the CO spatial template, 2FGL~J0631.6$+$0640 in Group S, and the 2FGL point sources except for the other two sources in Group S~{\bf(2FGL J0636.0$+$0554 and 2FGL J0637.8$+$0737)},
the parameters of which were the best-fit ones obtained by the likelihood analysis with the CO template plus {\bf all 3 point sources of Group S}~(Model 2 in Table~\ref{tab:likeratio1}).
A spatially extended emission region within the SNR becomes apparent.
%A diffuse-like emission emerged. 

\begin{figure}
%\epsscale{1.5}
%\plotone{figs/residual/residual.eps}
%\plotone{figs/residual/residual_v2_160122.eps}
\plotone{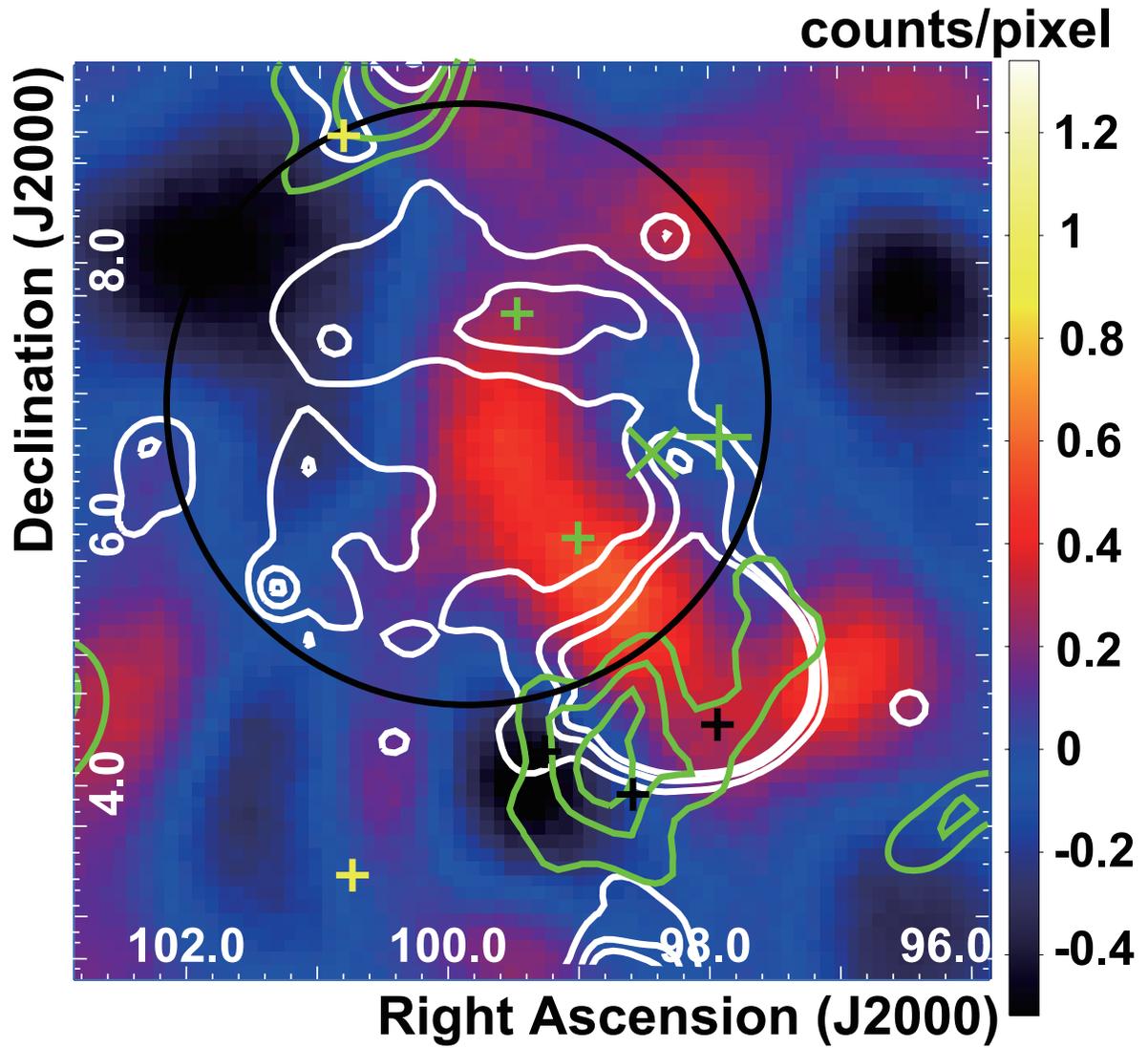}
\caption{
Background-subtracted LAT counts map in the 0.5--10~GeV energy range.
Here, the background model is the best-fit spatial model described in the text.
The details of the binning and smoothing are the same as Figure~\ref{fig:subtract_cmap2}.
The scale of the color bar is also the same as Figure~\ref{fig:subtract_cmap2}.
\label{fig:residual}
}
\end{figure}

To quantitatively evaluate the detection significance of the spatially extended emission,
we tested the hypothesis of an extended emission region inside the Monoceros Loop against the model with the two individual point sources in Group S~{\bf(2FGL J0636.0$+$0554 and 2FGL J0637.8$+$0737)}.
%from the 2FGL catalog. 
%We adopted a uniform disk or a Gaussian profile as a spatial template assuming a simple power-law spectrum.
We fitted the LAT counts by 
replacing {\bf all 3 point sources of Group S} by one point source~(2FGL~J0631.6$+$0640 in Group S) 
and a spatially extended component modeled as either a uniform disk or a Gaussian emission profile~(Model 3 and 4 in Table~\ref{tab:likeratio1}).
%plus one uniform disk model or a Gaussian profile model.
%where the radius of the disk is 1.$^\circ$73 corresponding to the radius of the radio shell,  
%and the position is the SNR center (R.A., Dec.)~$=$~(6${}^h$39${}^m$00${}^s$,6${}^\circ$30$'$00$''$).
%We note that it is not easy to use the radio emission as a spatial template because a strong emission from the Rosette Nebula is thermal.
%and the position of the disk are given by the shape of the radio emission from the SNR. 
%The position of the point source was set by eye to (R.A., Dec.)~$=$~(6${}^h$37${}^m$28${}^s$,6${}^\circ$01$'$48$''$).
The spectral shapes for both additional sources were assumed to be power-law functions.
We varied the radius~(1~$\sigma$ for a Gaussian profile) and location of the extended components and evaluated the maximum likelihood values.
The resulting maximum likelihood values with respect to the maximum likelihood for the null hypothesis are summarized in Table~\ref{tab:likeratio1}. 
The TS values for the uniform disk or the Gaussian profile plus {\bf the} point source and the CO template~(Model 3 and 4) are much larger than for the three 2FGL sources plus the CO template~(Model 2) albeit having fewer degrees of freedom.
%with less additional degrees of freedom.
%The TS value for the disk plus one point source with the CO template is much larger than for the three 2FGL sources $+$ the CO template with less additional degrees of freedom.
The Gaussian profile~(Model 4) provides a greater likelihood than the uniform disk~(Model 3): the TS value increases by 26.7.
%26.7 larger in TS value.
%In addition,
The maximum likelihood Gaussian profile has a radius~($\sigma$) of $2\fdg3^{+0.6}_{-0.5}$ centered on 
%(R.A., Dec.)~$=$~(6${}^h$39${}^m$25.62${}^s$, 6${}^\circ$55$'$43.1$''$)~(J2000).
(R.A., Dec.)~$=$~(99\fdg86,6\fdg93)~(J2000).
The error of the centroid is 0\fdg35 at 68\% confidence level.
The detection significances for the best-fit Gaussian profile and 2FGL~J0631.6$+$0640 
at energies of $>$~0.5~GeV are $\sim$~14~$\sigma$ and $\sim$~11~$\sigma$, respectively.
% Also, the diffuse emission was more evident after the subtraction of the emission from 2FGL~J0631.6$+$0640 as shown in Figure~\ref{fig:subtract_cmap3}.
We note that if we add the five eliminated 2FGL sources on top of the best-fit model, the TS value increases by only 6.2 
for 10 additional degrees of freedom, i.e. there is no statistical evidence for the presence of these sources in addition to the extended templates.
Also, we note that the maximum likelihood value for a 408~MHz radio template with suppression of emission from Rosette Nebula~(Model 5) was significantly worse than the best-fit Gaussian model, indicating that the gamma-ray emission around the Monoceros Loop is not strongly spatially associated with the shock region of the SNR as traced by radio.
Finally, we examined the residual map after fitting as shown in Figure~\ref{fig:residual}.
There is no prominent gamma-ray emission left in the map.
Therefore we adopted the Gaussian template with maximum likelihood parameters for the whole SNR in the following spectral analysis.

\subsection{Spectral analysis} 
\begin{figure}
%\epsscale{1.20}
%\plotone{figs/mono_spec/plotFlux_PowerLaw_Mono_v2.eps}
%\plotone{figs/spec_all/spec_all.eps}
\epsscale{1}
%\plotone{figs/spec_all/spec_all_v2.eps}
%\plotone{figs/mono_spec/spec_PlotFlux_LogParabola_sys_fit_line_150908.eps}
\plotone{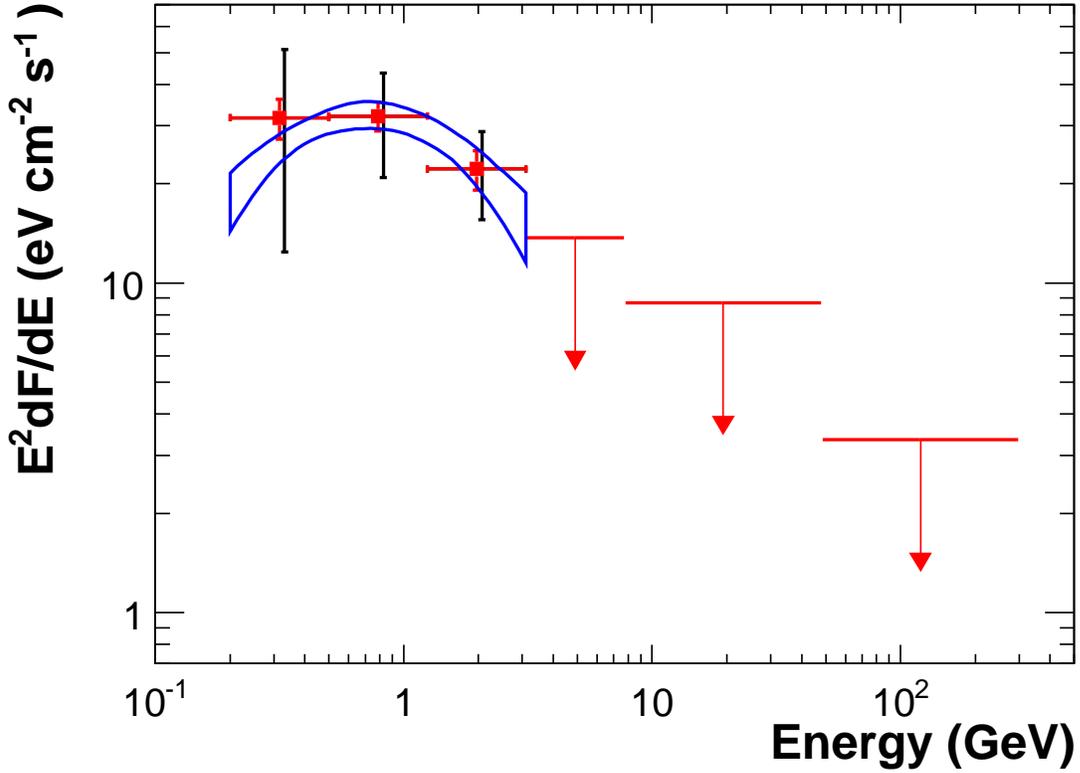}
\caption{ 
Spectral energy distribution of gamma-ray emission toward the Monoceros Loop.
The measured LAT fluxes are shown as red squares with horizontal lines indicating the energy range. Statistical and systematic error bars are shown in red and black, respectively. Flux upper limits at the 90\% confidence level are shown for energy bins when the detection was not significant (test statistic $<$~4) . The blue region is the 68\% confidence range~(no systematic error) of the LAT spectrum  assuming that the spectral shape is a log parabola.
 \label{fig:spec_all}}
\end{figure}

\begin{figure}
%\epsscale{1.20}
%\plotone{figs/spec_multi_rosette/spec_multi_rosette_v2.eps}
%\plotone{figs/spec_multi_rosette/spec_multi_rosette_v3.eps}
\plotone{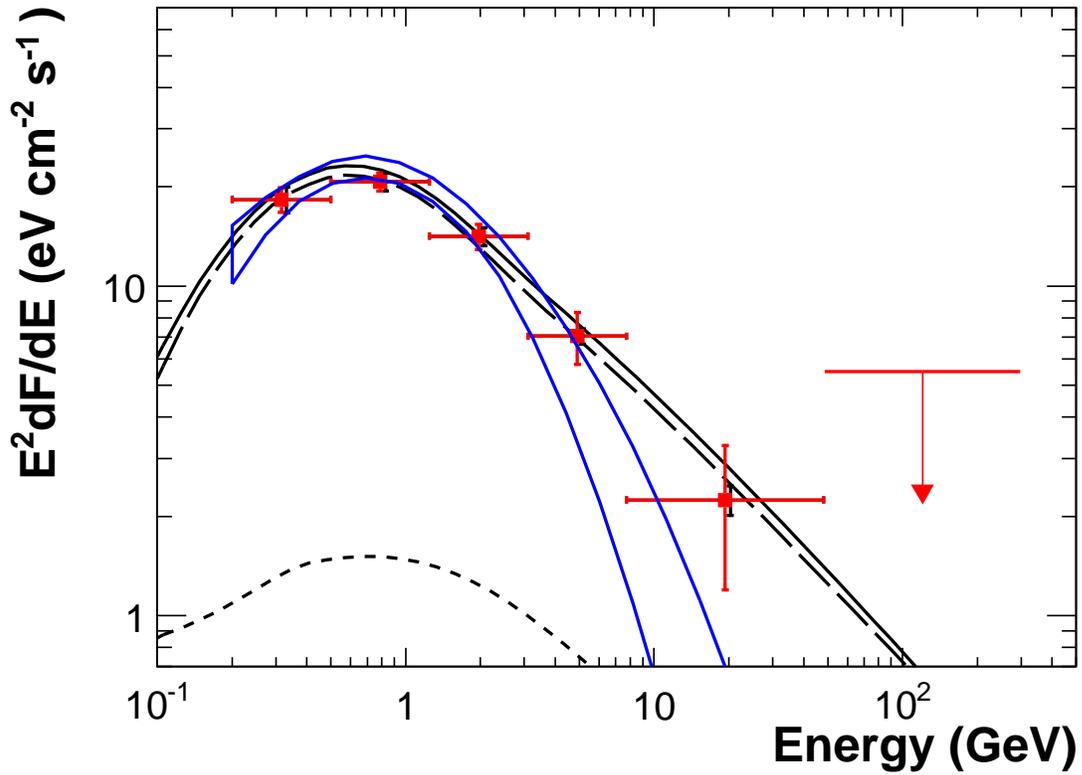}
\caption{ Spectral energy distribution of the gamma-ray emission measured by the LAT for the Rosette Nebula and the spectrum expected from the model discussed in the text.
The details of the LAT spectrum and the modeled emission processes are described in the captions of Figure~\ref{fig:spec_all} and \ref{fig:spec_multi}, respectively.
 \label{fig:spec_multi_rosette}}
\end{figure}

%To measure the spectra for the Rosette Nebula, the SNR, and the point source within the SNR, 
To measure the spectra of the SNR and the Rosette Nebula we {\bf used a maximum likelihood fit using the best-fit spatial model over the energy range from 0.2~GeV to 300~GeV}. Figures~\ref{fig:spec_all} and \ref{fig:spec_multi_rosette} show the resulting spectral energy distributions~(SEDs) for the SNR and the Rosette Nebula, respectively. {\bf If the detection is not significant in an energy bin, i.e., the improvement of the TS value with respect to the null hypothesis is less than 4~(corresponding to 2~$\sigma$ for one additional degree of freedom) then we calculated a 90\% confidence level upper limit assuming a photon index of 2.}

At least three different sources of systematic uncertainties affect our analysis: uncertainties in the LAT event selection efficiency, the adopted diffuse model and the morphological templates. Uncertainties in the LAT effective area were evaluated by comparing the efficiencies of analysis cuts for data and simulation of observations of Vela and the limb of the Earth, among other consistency checks~\citep{IRFs}. 
For \texttt{P7REP\_SOURCE\_V15}, these studies suggest 
a 10\% systematic uncertainty below 100~MeV, decreasing linearly with the logarithm of energy to 5\% in the range between 316~MeV and 10~GeV and increasing linearly with the logarithm of energy up to 15\% at 1~TeV. 
%estimated to 
%be 10\% at 100~MeV, decreasing to 5\% at 560~MeV, and increasing to 10\% at 10~GeV and the above, linearly varying with the logarithm of energy between those values
%We adopted the strategy described in \cite{diffusesys} to evaluate the systematic uncertainties due to the modeling of interstellar emission.
We adopted the strategy described in \cite{diffusesys1} and \cite{SNRCat} to evaluate the systematic uncertainties due to the modeling of interstellar emission. Results obtained using the standard model in Section~\ref{subsec:spatial} were compared with the results of eight alternative interstellar emission models. These models were created by varying the uniform spin temperature used to estimate the column densities of interstellar atomic hydrogen, the vertical height of the CR propagation halo, and the CR source distribution in the Galaxy.
%We adopted the strategy
%described in \cite{diffusesys} to evaluate
%the systematic uncertainties due to the modeling of interstellar emission.
%The difference, namely the local departure from the best-fit diffuse model, is found to be 6\%. By changing the normalization of the Galactic diffuse model artificially by $\pm$~6\%, we estimated the systematic errors.

We similarly gauged the uncertainties due to the morphological template by comparing the results with those obtained by changing the radius of the Gaussian template within its $\pm$~1~$\sigma$ error.
The total systematic errors were set by adding the above uncertainties in quadrature.
If the total systematic error in an energy bin was $>$~100\%, the point was replaced by an upper limit.
This is relevant for the fourth energy bin (3.105~GeV -- 7.746~GeV) in Figure~\ref{fig:spec_all} where an upper limit is presented due to the large systematic error although the TS value is $\sim$~9. 
The dominant systematic error for the measurement of the SNR spectrum arises from the uncertainty of the diffuse model below 0.5~GeV and the morphological uncertainty above 0.5~GeV, respectively.

We searched for a spectral break in the LAT energy range by comparing the likelihood values of a spectral fit over the whole energy range considered based on a simple power law and a log parabola function. TS values and best-fit parameters are summarized in Table~\ref{tab:spectral_shape}. The values for a log parabola function correspond to improvements at the $> 6$~$\sigma$ confidence level for the SNR and $> 9$~$\sigma$ for the Rosette Nebula when only statistical uncertainties are taken into account.
For the SNR, we further investigated the systematic effects on the above spectral analysis. 
Accounting for systematics in the fit, the curved shape is still preferred over a power-law at a confidence level $>$~5~$\sigma$.
%A log parabola function is the best for both sources.
By comparing the spectral parameters of a log parabola function for both sources, the spectral shapes are consistent within the statistical errors at our current sensitivity.
Assuming the spectral shape is a log parabola function, 
the gamma-ray luminosities 
integrated over the energy range 0.2--300~GeV
 inferred from our analysis are $\sim$~$4 \times 10^{34}$~erg~s$^{-1}$ for the SNR and $\sim$~$3 \times 10^{34}$~erg~s$^{-1}$ for the Rosette Nebula, respectively.

\section{DISCUSSION}
\label{sec:discuss}
%We find an extended region of gamma-ray emission spatially coincident with the Monoceros SNR. 
An extended region of gamma-ray emission was found to be spatially coincident with the Monoceros SNR by \cite{SNRCat}. We confirmed the extended emission with this more detailed analysis.
Since no pulsar wind nebula has been discovered so far within the SNR~\citep[e.g.,][]{PWNcatalog}, the likely explanation for the bulk of this gamma-ray emission is the interaction of high-energy particles accelerated in the shocks of the Monoceros Loop with ambient interstellar matter and radiation fields. 
The morphological difference between the gamma-ray emission and the radio emission can be explained by the inhomogeneity of the nearby gas, which is irradiated by the accelerated CRs that have escaped from the shocked regions.
This hypothesis would also readily explain the enhanced emission from the nearby Rosette Nebula where the same population of high-energy particles would produce a bright gamma-ray signal when interacting in the dense molecular clouds traced by the CO emission. 
We note that the possibility of a pulsar wind nebula without detectable radio emission cannot be ruled out for the explanation of the enhanced emission around the SNR.
Also, we cannot rule out that some of the emission around the SNR is produced by dark gas, i.e. gas that is not accounted for in HI or CO surveys. Its distribution cannot be modeled precisely, yet large quantities of dark gas have been found surrounding nearby molecular clouds~\citep{darkgas}. In contrast, it is difficult to explain the enhanced emission around the Rosette Nebula only by dark gas considering a good fit of the CO template and the feature of dark gas that is mostly at the outskirt of the cloud.
%There seems to be a correspondence between gamma-ray emission and the SNR. 
%around the SNR and radio continuum emission originated by high-energy electrons via synchrotron radiation,
%indicating that the high-energy particles responsible for gamma-ray emission are also accelerated in the SNR.
%are in the vicinity of the shock regions.
%If so, enhancement of gamma-ray emission from the Rosette Nebula, which is the nearby molecular clouds traced by CO line emission, can be reasonably explained by interaction between high-energy particles and the dense material in the clouds.
%Thus we argue that the bulk of gamma-ray emission comes from interactions of high-energy particles accelerated at the shocks of the Monoceros Loop with interstellar matter or fields in the regions.
%There are some possibilities of the origin of the emission from 2FGL~J0631.6$+$0640 within the SNR modeled by a point-like source. This source is not spatially associated with HESS~J0632$+$057 from the angular separation $\sim$ 0.$^\circ$9. 
%One possibility is the emission from a pulsar wind nebula due to the proximity to the gamma-ray pulsar.
%We do not focus on the origin of the emission from the point source here.
%We note that the existence of a pulsar wind nebula around the SNR has not been reported so far~\citep[e.g.,][]{PWNcatalog}.
%Also, 
%We note that we cannot rule out the possibility of emission from dark gas along the line of sight due to the imperfections in the modeling of dark gas in the diffuse model.

{\bf Broadband emission from the Monoceros Loop SNR was modeled under the assumption that gamma rays are emitted by a population of accelerated protons and electrons. We assumed relativistic electrons and protons have the same injection spectrum and occupy the same spatial volume characterized by a constant magnetic field strength and matter density. We used the following equation to model the momentum distribution of injected particles: }
\begin{equation}
Q_{e, p}(p) = a_{e,p} \left( \frac{p}{1~{\rm GeV}~c^{-1}} \right)^{-s_{\rm L}} \left\{ 1+\left(\frac{p}{p_{\rm br}} \right)^2 \right\} ^{-(s_{\rm H}-s_{\rm L})/2},
\end{equation}
 where $p_{\rm br}$ is the break momentum, $s_{\rm L}$ is the spectral index below the break
and $s_{\rm H}$ above the break. 
{\bf $a_{e,p}$ are normalizations for the electron and proton components, respectively.}
{\bf Because the details of the proton/electron injection process are poorly known, we adopt a minimum momentum of 100~MeV~$c^{-1}$ .}

Electrons suffer energy losses due to ionization, Coulomb scattering, bremsstrahlung, synchrotron
emission and inverse Compton~(IC) scattering. 
%The evolution of the electron momenta spectrum 
{\bf The evolution of the momenta spectra $N_{e,p}(p,t)$ are}
 calculated from the following equation:
\begin{equation}
  \frac{\partial N_{e,p}}{\partial t} = \frac{\partial}{\partial p} \left( b_{e,p}N_{e,p} \right) + Q_{e,p} ,
\end{equation}
where $b_{e,p} = -dp/dt$ is the momentum loss rate, and $Q_{e,p}$ is the particle injection rate.
{\bf We assumed that the shock produced particles at a constant rate, so $Q_{e,p}$ is constant. }
%This prescription approximates the weakening of the shock and the reduction in the particle acceleration efficiency.} 
% until the SNR enters the radiative phase, at which time the source turns off
 To derive the gamma-ray emission spectrum we calculated $N_{e,p}(p,T_{\rm 0})$
numerically, where $T_{\rm 0}$ is the SNR age of $3 \times 10^4$~yr. {\bf Momentum losses for protons are neglected because the timescale for radiative losses via neutral pion production is} $\sim$~10$^7/(\nh/1~{\rm cm^{-3}})$~yr where $\nh$ is the gas density averaged over {\bf the volume occupied by high-energy particles}. {\bf Gamma-ray emission by secondary leptons produced from charged pion decay was neglected. Generally, this is a negligible contribution unless the gas density is comparable to that in dense molecular clouds and the SNR has reached the later stages of its evolution, or the injected electron-to-proton ratio is much lower than locally observed. The calculation of the spectrum of $\pi^0$ decay gamma rays from interactions between protons and ambient hydrogen was adopted from \cite{Dermer86}. A scaling factor of 1.84 accounted for helium and heavier nuclei in target material and CRs \citep{Mori09}.} Contributions from bremsstrahlung and IC scattering by accelerated electrons are computed based on \cite{Blumenthal70}, and synchrotron radiation is evaluated using the work of \cite{Crusius86}.

\begin{figure}
%\plotone{./figs/spec_multi/spec_multi.eps}
%\plotone{./figs/spec_multi/spec_multi_v2.eps}
% \plotone{./figs/spec_multi/spec_multi_v3.eps}
%\plotone{./figs/spec_multi/spec_multi_150910.eps}
\plotone{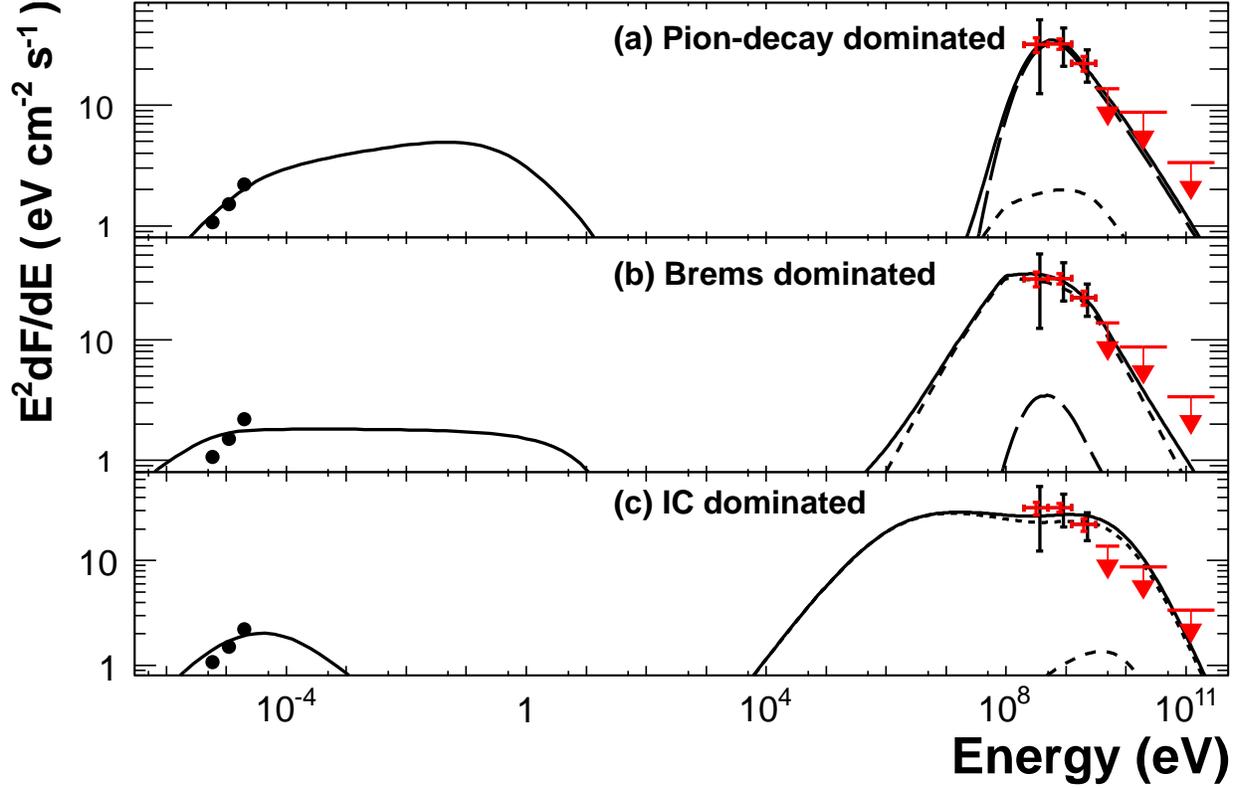}
\caption{Multi-band spectrum of the Monoceros Loop. \label{fig:spec_multi}
{\bf
LAT measurements reported in Figure~\ref{fig:spec_all} are shown alongside radio continuum measurements~\citep{RadioSpectrum}. Radio emission is modeled as synchrotron radiation, 
while gamma-ray emission is modeled by different combinations of $\pi^0$-decay~(long-dashed
curve), bremsstrahlung~(dashed curve), and inverse Compton~(IC) scattering~(dotted curve). 
As described in the text, the models are: a) $\pi^0$-decay dominated, b)
bremsstrahlung dominated, c) IC-dominated.
}
%, d) IC-dominated model with 100 times higher photon field than c).
}
\end{figure}

First, we considered a model with the Monoceros Loop SNR dominated by $\pi^0$-decay.
The gamma-ray spectrum constrains the number index of accelerated protons to be $s_{\rm H} \approx 2.8$ in the high-energy regime. We adopted a spectral index $s_{\rm L} = 1.5$ to explain the radio continuum spectrum~\citep{RadioSpectrum}. We note that the radio spectrum was estimated from the full SNR with the exception of the Rosette Nebula region that is dominated by strong thermal emission.
Since we expect curvature in the GeV energy band due to the kinematics of $\pi^0$ production and decay, it is difficult to constrain a break in the proton momenta spectrum from the gamma-ray spectrum. The gamma-ray spectrum thus provides only an upper bound to the momentum break at $\sim$~10~GeV~$c^{-1}$. We adopt a break at the best-fit value, 2~GeV~$c^{-1}$.
The density is fixed to $3.6~\mathrm{cm}^{-3}$ based on the \hi observations~\citep{RadioSpectrum}.
The resulting total proton energy,  $W_{p}\sim 7.6~\times~10^{49}~\cdot~(3.6~\mathrm{cm}^{-3}/\nh)\cdot(d/1.6 \mathrm{kpc})^{2}$~erg, is less than 10\% of the typical kinetic energy of a supernova explosion. {\bf For the electron-to-proton ratio measured at Earth, $K_{ep} \equiv a_e/a_p = 0.01$, } the magnetic field strength is determined to be $B \sim 35\;\upmu$G by {\bf the} radio data. Using these model parameters (Table~\ref{tab:model}), we obtained the SEDs shown in Figure~\ref{fig:spec_multi}~(a).

In the case of leptonic scenarios, we assume $K_{ep} = 1$ to produce the gamma-ray emission predominantly from the electrons. The radio spectrum~\citep{RadioSpectrum} is difficult to {\bf be modeled as the synchrotron radiation when we fit the gamma-ray spectrum with } a model dominated by electron bremsstrahlung, as shown in Figure~\ref{fig:spec_multi}~(b).

{\bf The other leptonic scenario is an IC-dominated model.}
IC gamma rays originate from the interaction of high-energy electrons with the cosmic microwave background~(CMB) as well as optical and infrared radiation fields. {\bf Galactic radiation fields were adopted from ~\citet{Porter08} at the location of the Monoceros Loop. These very complex spectra are approximated by two infrared and two optical blackbody components.}
It is hard to reproduce the multi-wavelength spectrum well with an IC-dominated model shown in Figure~\ref{fig:spec_multi}~(c).
 {\bf In addition, the ratio between IC and synchrotron fluxes constrained} the magnetic field to be less than $\sim$~2~$\upmu$G
and requires a low gas density of $\nh \sim 0.01~\mathrm{cm}^{-3}$ to suppress the electron bremsstrahlung, which is unlikely.

In conclusion, the bulk of the gamma-ray emission from the Monoceros SNR is most likely from $\pi^0$ decay produced by the interactions of protons with ambient hydrogen. It is then reasonable to explain the gamma-ray spectrum of the Rosette Nebula by the same process.
%Assuming that $\pi^0$-decay produced by the interactions of protons with ambient hydrogen is responsible for the bulk of the gamma-ray emission from the SNR,the gamma-ray spectrum of the Rosette Nebula can be reasonably explained by a $\pi^0$-decay dominated model.
If the protons are accelerated in the whole SNR in the same manner and are not strongly affected by spectral deformation due to CR diffusion processes,
the shape of the proton spectrum in the Rosette Nebula is expected to be the same as in the Monoceros Loop.
%the gamma-ray spectrum of Rosette Nebula is expected to be modeled with the same shape of a proton momentum spectrum.
Figure~\ref{fig:spec_multi_rosette} shows the gamma-ray spectrum of the Rosette Nebula with the $\pi^0$-decay dominated model assuming the density in the molecular clouds is $100~\mathrm{cm}^{-3}$. 
The spectrum can be reproduced without any change from
the proton momentum spectrum of the SNR.
% {\bf The $s_{\rm H}$ is preffered to be a softer value 2.8, although it is not strictly constrained due to the flux errors.}
The resulting total proton energy,  $W_{p}\sim 0.18~\times~10^{49}~\cdot~(100~\mathrm{cm}^{-3}/\nh)\cdot(d/1.6 \mathrm{kpc})^{2}$~erg, is about 2\% 
of that for the $\pi^0$-decay model of the SNR, which is reasonable considering the solid angle of the Rosette Nebula with respect to the SNR and uncertainty of the matter density.
%The goodness of the fit supports our assumption.
We note that these CR energies for the Monoceros Loop SNR and the Rosette Nebula are the enhancements of the CR density in addition to that implicit in the standard Galactic diffuse emission model.

To summarize, 
the assumption that the gamma-ray emission from the Monoceros SNR is dominated
%it is most natural to assume that gamma-ray emission from the Monoceros Loop is dominated 
by decay of $\pi^0$ produced in nucleon-nucleon interactions of hadronic CRs with interstellar matter
is a natural scenario that can
also readily explain the emission from the nearby Rosette Nebula as interactions of the same population of CRs in the dense molecular cloud.
Similarly to SNR HB~3~\citep{HB3},
it should be emphasized that our observations towards the Monoceros Loop provide a rare and valuable example for which the emissions from both the SNR and the interacting molecular clouds are detected. 
%A similar example has been recently reported for SNR HB~3~\citep{HB3}.
%It should be emphasized that our observations of the Monoceros Loop combined with the radio data constrain the proton momentum break to be below 10~GeV~$c^{-1}$.
%in the range, 2--10~GeV~$c^{-1}$.

\section{CONCLUSIONS}
\label{sec:conclusion}

We analyzed gamma-ray measurements by the LAT in the region of the Monoceros Loop.
%, detecting significant gamma-ray emission spatially associated with the remnant and the Rosette Nebula including molecular clouds.
%There is a correspondence between gamma-ray emission around the SNR and radio continuum emission originated by high-energy electrons via synchrotron radiation,indicating that the high-energy particles responsible for gamma-ray emission are in the vicinity of the shock regions.
The brightest gamma-ray peak is spatially correlated with the Rosette Nebula.
%which was well reproduced by a CO spatial template. 
A template derived from the CO gas distribution fits the morphology of the gamma-ray emission better than a set of individual point sources.
%After subtraction of this emission by spatial modeling, the gamma-ray emission from the SNR emerged, which was reproduced by a uniform-disk spatial template.
Gamma-ray emission from an extended source was also found inside the Monoceros Loop. 
%An emission profile corresponding to a Gaussian profile
A Gaussian emission profile of $2\fdg3$ radius~(1~$\sigma$) is a substantially better match to the gamma-ray data when compared to the 2FGL sources
J0636.0$+$0554 and J0637.8$+$0737.
The gamma-ray spectra of both extended components were
 significantly better reproduced by a curved shape than a simple power law.

Their respective luminosities
integrated over the energy range 0.2--300~GeV are 
%$\sim$~$4.1 \times 10^{34}$~erg~s$^{-1}$ for the SNR and $\sim$~$2.7 \times 10^{34}$~erg~s$^{-1}$ for the Rosette Nebula. 
$\sim$~$4 \times 10^{34}$~erg~s$^{-1}$ for the SNR and $\sim$~$3 \times 10^{34}$~erg~s$^{-1}$ for the Rosette Nebula. 
The decay of $\pi^0$ produced by interactions of hadrons accelerated
by the remnant with interstellar gas can naturally explain the gamma-ray emission of the remnant and that of the nebula.
%, although the other possibilities are not completely ruled out.
%In this scenario our observations of the Monoceros Loop indicate that the proton momentum spectrum is steep in the high-energy regime, with a spectral break which is constrained together with radio continuum emission below 10~GeV~$c^{-1}$. 
%in the range 2--10~GeV~$c^{-1}$.

\acknowledgments
% added on Mar 23, 2014
The \textit{Fermi} LAT Collaboration acknowledges generous ongoing support
from a number of agencies and institutes that have supported both the
development and the operation of the LAT as well as scientific data analysis.
These include the National Aeronautics and Space Administration and the
Department of Energy in the United States, the Commissariat \`a l'Energie Atomique
and the Centre National de la Recherche Scientifique / Institut National de Physique
Nucl\'eaire et de Physique des Particules in France, the Agenzia Spaziale Italiana
and the Istituto Nazionale di Fisica Nucleare in Italy, the Ministry of Education,
Culture, Sports, Science and Technology (MEXT), High Energy Accelerator Research
Organization (KEK) and Japan Aerospace Exploration Agency (JAXA) in Japan, and
the K.~A.~Wallenberg Foundation, the Swedish Research Council and the
Swedish National Space Board in Sweden.
 
Additional support for science analysis during the operations phase is gratefully
acknowledged from the Istituto Nazionale di Astrofisica in Italy and the Centre National d'\'Etudes Spatiales in France.

We thank Luigi Tibaldo for helpful comments and discussions on the dark gas.

%\begin{figure}
%\plotone{figs/map2_v2/monoceros_ps.eps}
%\caption{
%Background-subtracted LAT counts map in the 0.5--10~GeV energy range.
%The 2FGL point sources except for 3 sources associated with the SNR~(2FGL~J0631.6$+$0640, 2FGL~J0636.0$+$0554, 2FGL~J0637.8$+$0737)
%are not included in the background model. 
%The counts map is binned using a grid of $0.^\circ1$ and smoothed with a Gaussian kernel of $\sigma$ $=$0.$^\circ$5.
%The details of the overlays are described in the caption of Figure~\ref{fig:subtract_cmap}.
%\label{fig:subtract_cmap2}
%}
%\end{figure}

%\begin{figure}
%\plotone{figs/rosette_spec/plotFlux_PowerLaw_Rosette_v2.eps}
%\caption{ Spectral energy distribution of the gamma-ray emission measured by the LAT for the Rosette Nebula. 
%The details of the plots and bars are described in the caption of Figure~\ref{fig:spec_mono}.
% \label{fig:spec_rosette}}
%\end{figure}

\begin{table}
\begin{center}
\caption{ Test Statistics for different spatial models compared to the null hypothesis of a model with no source associated
with the Rosette Nebula and the Monoceros Loop~(0.5--300~GeV) \label{tab:likeratio1}}
\begin{tabular}{lccc}
\tableline\tableline
 Model   & Test Statistic\tablenotemark{a} &  Additional Degrees of  \\
         &                                 &  Freedom  \\
\tableline
   1: 3 point sources of Group R $+$ 3 point sources of Group S\tablenotemark{b} &  424.1 &  12 \\
   2: CO image + 3 point sources {of Group S} & 728.8 & 9\tablenotemark{c}  \\
%   CO image + Uniform disk $+$ 2FGL~J0631.6$+$0640  & 916.4 & 7 \\
   3: CO image + Uniform disk $+$ 2FGL~J0631.6$+$0640\tablenotemark{d}  &  937.6 & 10 \\
   4: CO image + Gaussian $+$ 2FGL~J0631.6$+$0640  & 964.3 & 10 \\
   5: CO image + Radio template\tablenotemark{e} $+$ 2FGL~J0631.6$+$0640  &  922.2 & 7 \\
\tableline
\end{tabular}
\tablenotetext{\rm a}{$-2\ln (L_{\rm 0}/L)$,
 where $L$ and $L_{\rm 0}$ are the maximum likelihoods for the model with/without the source component, respectively. The model for $L_{\rm 0}$ includes PSR~J0633$+$0632.}
\tablenotetext{\rm b}{
{\bf The 3 sources in the 2FGL source list associated with the Rosette Nebula~\citep{2yrCatalog} are referred to as Group R in the text. The 3 sources listed in the 2FGL source list associated with the Monoceros Loop are referred to as Group S in the text.}
%The six sources listed in the 2FGL source list associated with the Rosette Nebula and the Monoceros Loop~\citep{2yrCatalog}.
2FGL~J0633.7$+$0633~(PSR~J0633$+$0632) is not included in them.
}
%\tablenotetext{\rm c}{The three sources listed in the 2FGL source list associated with the Monoceros Loop~\citep{2yrCatalog}.
%2FGL~J0633.7$+$0633~(PSR~J0633$+$0632) is not included in them.
%}
\tablenotetext{\rm c}{The additional degrees of freedom for the CO image is 2 for the spectral shape, 1 for the analysis threshold to extract emission. The details are shown in the text.}
\tablenotetext{\rm d}{{\bf 2FGL~J0631.6$+$0640 is included in Group S.}}
\tablenotetext{\rm e}{The radio template was obtained from 408~MHz radio data~\citep{RadioTemplate} by excluding the region around the Rosette Nebula, where the emission is predominantly thermal.
The additional degrees of freedom for the radio template are 2 for the spectral shape~(a power law).}

\end{center}
\end{table}

\begin{table}
\begin{center}
\caption{Test Statistics and Parameters for Spectral Models~(0.2--300~GeV) \label{tab:spectral_shape}}
\begin{tabular}{lcccc}
\tableline\tableline
 Spectral Model  & Test Statistic\tablenotemark{a}  &  Additional Degrees of & Spectral Parameters  \\
  &  &    Freedom & \\
\tableline\tableline
Monoceros Loop &   &  &  \\
\tableline
Power Law  & 0  & 2 & $E^{-p}$; $p=2.27\pm0.03$ \\
\tableline
Log Parabola  & 51 &  3 &   $\left(\frac{E}{1~{\rm GeV}}\right)^{-p_1-p_2\log{\left(\frac{E}{1~{\rm GeV}}\right)}}$ \\
 & & &  $p_1=2.23 \pm 0.06$ \\
 & & &  $p_2=0.35 \pm 0.03$ \\
%\tableline
%Smoothly broken power law  & 54 & 4  & $E^{-p_1}\left\{ 1+\left(\frac{E}{E_{\r%m b}}\right)^{\frac{-p_1+p_2}{0.2}} \right\}^{-0.2}$    \\
% & & &  $p_1= -0.14 \pm 0.33$ \\
% & & &  $p_2= 2.49 \pm 0.07$ \\
% & & &  $E_{\rm b}= 0.38 \pm 0.03$~GeV \\
\tableline\tableline
Rosette Nebula &   &  &  \\
\tableline
Power Law  & 0  & 2 & $E^{-p}$; $p=2.32\pm0.02$ \\
\tableline
Log Parabola  & 88 &  3 &   $\left(\frac{E}{1~{\rm GeV}}\right)^{-p_1-p_2\log{\left(\frac{E}{1~{\rm GeV}}\right)}}$ \\
 & & &  $p_1=2.29 \pm 0.03$ \\
 & & &  $p_2=0.39 \pm 0.02$ \\
%\tableline
%Smoothly broken power law  & 94 & 4  & $E^{-p_1}\left\{ 1+\left(\frac{E}{E_{\rm b}}\right)^{\frac{-p_1+p_2}{0.2}} \right\}^{-0.2}$    \\
% & & &  $p_1= 1.04 \pm 0.31$ \\
% & & &  $p_2= 2.72 \pm 0.09$ \\
% & & &  $E_{\rm b}= 0.58 \pm 0.09$~GeV \\
\tableline
\tableline\tableline
\end{tabular}
\tablenotetext{\rm a}{$-2\ln (L_{\rm 0}/L)$,
 where $L$ and $L_{\rm 0}$ are the maximum likelihood values for the model under consideration and the power-law model, respectively.}
%\tablecomments{The test statistics for the best-fit uniform ring with exponential cutoff, log parabola, and smoothly broken power law with respect to the null hypothesis of no emission associated with the Cygnus Loop are 572, 580, and 581 in the energy band 0.2--100~GeV.
%, corresponding to a detection significance of $\sim$~23~$\sigma$.}
\end{center}
\end{table}

\begin{table}
\begin{center}
\caption{Model parameters for the Monoceros Loop.\label{tab:model}}
\begin{tabular}{lccccccccc}
\tableline\tableline
 Model  & \kep\tablenotemark{a} &  $s_{\rm L}$\tablenotemark{b} & $p_{\rm br}$\tablenotemark{c} & $s_{\rm H}$\tablenotemark{d} & $B$ &
 $\nh$\tablenotemark{e} & $W_{p}$\tablenotemark{f} & $W_{e}$\tablenotemark{f} \\
   &    &  & (GeV~$c^{-1}$) & &  ($\upmu$G) & (cm$^{-3}$) & ($10^{49}$~erg) & ($10^{49}$~erg) \\
\tableline
(a)~Pion & 0.01  & 1.5 & 2.0 & 2.8 & 35 & 3.6 & 7.6 & 0.087 \\
(b)~Bremsstrahlung & 1 & 1.5 & 1.0 & 3.0 & 9 & 3.6  & 0.98 & 1.4  \\
(c)~Inverse Compton\tablenotemark{g} & 1 & 1.5 & 20 & 4.0 & 1.1 & 0.01 & 12 & 16 \\
%(d)~Inverse Compton & 10 & 1.5 & 20.0 & 4.0 & 113 & 3.6 & 1.2~$\times 10^{-3}$ & 0.012 \\
%~~(photon field $\times$ 1000) & & & & & & & & \\
\tableline
\end{tabular}
\tablenotetext{\rm a}{The ratio electrons-to-protons
at 1~GeV~$c^{-1}$.}
\tablenotetext{\rm b}{
The momentum distribution of particles
 is assumed to be a smoothly broken power-law, where the indices and the break
 momentum are identical for both accelerated protons and electrons.
$s_{\rm L}$ is the spectral index in momentum below the break.}
\tablenotetext{\rm c}{$p_{\rm br}$ is the break momentum.}
\tablenotetext{\rm d}{Spectral index in momentum above the break.}
%\tablenotetext{\rm e}{The exponential cutoff momentum of particle distributions.}
\tablenotetext{\rm e}{Average hydrogen number density of ambient medium.}
\tablenotetext{\rm f}{The distance from the Earth is assumed to be 1.6~$\mathrm{kpc}$~\citep{distance1, distance2}. The total energy
 is calculated for particles $>$~100~MeV~$c^{-1}$.}
\tablenotetext{\rm g}{Seed photons for inverse Compton scattering of
 electrons include the CMB,
 two infrared~($T_{\rm IR} = 33.3, 4.95 \times 10^2$~K, $U_{\rm IR} = 0.20, 4.10 \times 10^{-2}$~eV~cm$^{-3}$, respectively), and
 two optical components~($T_{\rm opt} = 3.72 \times 10^3, 1.11 \times 10^4$~K, $U_{\rm opt} = 0.30, 0.11$~eV~cm$^{-3}$, respectively) in the vicinity of the Monoceros Loop.
}
\end{center}
\end{table}

\end{document}